\documentstyle[aps,epsf,eqsecnum,prb,graphicx,twocolumn]{revtex}  

\def\figwidth{8cm}

\def\iomn{i\omega_n}
\def\ep{\epsilon_{\perp}}

\def\tp{t_{\perp}}
\def\kp{k_{\perp}}

\def\sp{\sigma_{\perp}}

\def\o{\omega}
\def\R{\mbox{Re}}

\def\est{E^*}

\begin{document}



\title{Quasi-one-dimensional
organic conductors: dimensional crossover and some puzzles}

\author{S. Biermann$^{(1,2)}$, A. Georges
$^{(2,1)}$, T. Giamarchi$^{(1,2)}$ and A. Lichtenstein$^{(3)}$}

\address{
$^1$ Laboratoire de Physique
des Solides, CNRS-UMR 8502, UPS B\^at. 510, 91405 Orsay France
\\
$^2$ LPTENS CNRS UMR 8549 24, Rue
Lhomond 75231 Paris Cedex 05, France
\\
$^3$ University of Nijmegen, NL-6525 ED Nijmegen, The Netherlands
}


\maketitle

\section{Introduction and scope of the paper}

The nature of the metallic phase of interacting electron systems
depends strongly on dimensionality. The interplay between
interactions and dimensionality is an important issue for a large
number of materials, ranging from cuprate superconductors to
low-dimensional organic conductors \cite{jerome_revue_1d}. For
quasi-one dimensional conductors (such as the Bechgaard salts, which
will be the main subject of this paper), these issues become
crucial. Indeed in three dimensions, Fermi liquid (FL) theory
applies, whereas in one dimension a different kind of low-energy
fixed point known as a Luttinger liquid (LL) is found, with
physical properties quite different from that of a FL.
By varying the
anisotropy of the system, or the energy scale at which it is
probed, one can thus expect drastic changes in the physical
properties.

This is even more true when the filling
of the system is commensurate. In that case, interactions can lead
to an insulating behavior via the Mott transition. This phenomenon
occurs in all dimensions but the one-dimensional case is
particularly favorable \cite{voit_bosonization_revue}. In quasi
one-dimensional (Q1D) systems, interchain hopping can induce a
(deconfinement) transition from the Mott insulating (MI) state to
a metallic state, and crossovers between different metallic
behaviors. Understanding how such a deconfinement transition can
take place and what are the properties of the metallic phases is a
particularly challenging problem, for reasons explained below.

This paper is organized as follows:
\begin{itemize}
\item In Sec.~\ref{sec:organics_phys}, {\it some}
physical properties of the TMTSF and TMTTF organic compounds
will be briefly reviewed
\cite{jerome_revue_1d,jerome_organic_review,bourbonnais_jerome_review},
with an emphasis on those related to the above issues. As described
there, these compounds are three dimensional stacks of quarter- filled
chains, which makes them wonderful laboratories, in which all the questions
above can be addressed. Along the way, we shall point out some
open questions and puzzles associated with the physics of these
materials.

\item In Sec.~\ref{sec:modelling}, we turn to theoretical models, and
explain why {\it non-perturbative} methods are required to deal
with interchain hopping. We shall review an extension of the
dynamical mean-field approach
\cite{arrigoni_tperp_resummation,arrigoni_tperp_resummation_prb,%
georges_organics_dinfiplusone}
designed to deal with this problem (chain-DMFT), and describe its
recent application to coupled Hubbard chains
\cite{biermann_dmft1d_hubbard_short} and its bearing on the above
issues.
\end{itemize}

\section{Quasi one-dimensional organic conductors: some physical properties}
\label{sec:organics_phys}

\begin{figure}
\centerline{\includegraphics[width=\figwidth]{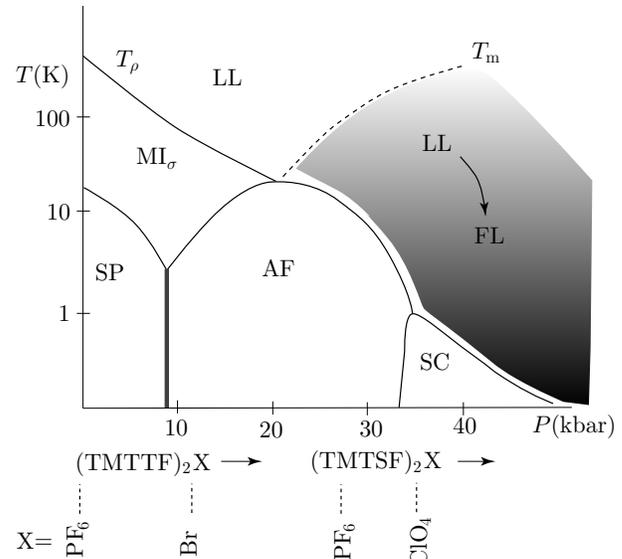}}
\caption{Unified experimental phase diagram for the TM compounds
(from \cite{bourbonnais_jerome_review}).
Either pressure or chemical changes (increasing pressure corresponds 
to going from the
TMTTF to the TMTSF family and changing the anions) yields the same 
phases [MI: Mott insulator,
LL: Luttinger liquid metal, FL: Fermi liquid metal,
SP: spin-Peierls, AF: antiferromagnetic spin-density wave, 
SC: superconducting].
The TMTTF family is insulating at ambient pressure whereas the TMTSF family
shows good metallic behavior at room temperature.}
\label{fig:unified}
\end{figure}

The Bechgaard salts TMTSF$_2$X were the first organic compounds to
exhibit superconductivity, and have thus been the focus of intense
experimental and theoretical studies. In addition to the superconducting
phase, these materials have a remarkably rich phase diagram
(cf. Fig.~\ref{fig:unified}) and exhibit a host of remarkable properties
(e.g non-FL metallic behaviour, quantized
Hall conductance, Fr{\"o}hlich conductivity), many of which are still
poorly understood.
Reviewing all these properties goes far beyond the scope
of this lecture: we shall restrict ourselves to a discussion of the
crossovers observed in the high-temperature regime, above the ordered
phases. For more extensive reviews of the physical properties of
quasi one-dimensional organics, see e.g
\cite{jerome_organic_review,bourbonnais_jerome_review}.

\subsection{Structure} \label{sec:salts}

\begin{figure}
\centerline{\includegraphics[width=\figwidth]{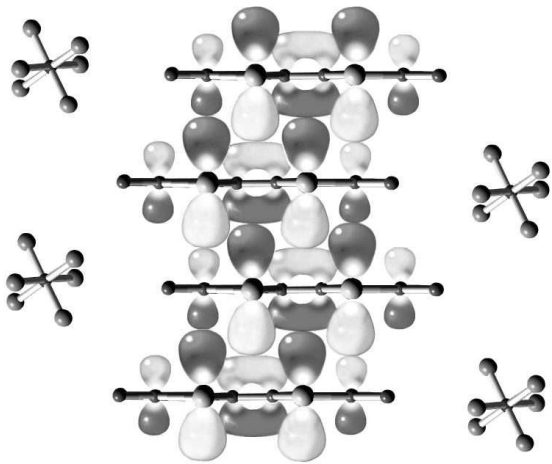}}

\caption{Structure of the Bechgaard salts (TMTSF)$_2$X 
(after \cite{bourbonnais_jerome_review}).
Electronic transport takes place preferentially along 
the stacks of TMTSF molecules (vertical, a-axis).
The horizontal axis (c-axis) for which the stacks are separated 
by the counter-ions is the least conducting one.}
\label{fig:bechgaard}\end{figure}

The basic building block of the Bechgaard salts
(Fig.~\ref{fig:bechgaard}) is the flat molecule
TMTSF (tetramethyltetraselenafulvalene), which contains four selenium atoms,
surrounded by four methyl groups in a fulvalene type double ring structure.
Also shown in Fig.~\ref{fig:bechgaard} are the orbitals
giving rise to the double bonds between the carbon atoms
and the $\pi$ orbitals of the selenium atoms.
In the TMTTF compounds (Fabre salts), the selenium atoms are replaced by
sulphur ones.

All these compounds crystallize in stacks of TM 
(short for TMTSF or TMTTF) molecules separated
by the counter-ions X (for example PF$_6$ or ClO$_4$).
The counter-ions are here to provide the charge neutrality, 
as in a standard salt.
The transfer of charge between the ion and the TM stacks 
is total. Since there is one ion
for two TM molecules, the chains are quarter filled. 
In addition the chains are slightly dimerized.
This raises the important question whether these systems 
should be considered as half-filled, rather
than quarter-filled, to which we shall come back 
in section~\ref{sec:transport}.
It is important to note that the commensurate 
filling is fixed by the chemistry
of the compound, and so far it has not been possible 
to move away from such a commensurate filling.
No doubt that if this could be done (through e.g. 
a field effect transistor
geometry \cite{schon_gate_superconductivity_polymer}), 
this would prove very interesting.

The overlap of the $\pi$- orbitals of the selenium or
sulfur atoms leads to a high mobility of electrons
along the stacking direction; the hopping integrals
in the perpendicular directions are indeed smaller
by more than one order of magnitude.
Estimated values of the hopping integrals along
the stack direction (a-axis) and the two perpendicular axes
pointing towards neighboring stacks (b-axis) and towards
the anions (c-axis) respectively are:
$t_a:t_b:t_c = 1000K:100K:30K$.
Therefore one can think of these materials as
one-dimensional chains coupled by small inter-chain
hoppings. Given the hierarchy of transverse coupling 
the system is first expected to
become two dimensional and then three dimensional at low temperatures.
At very low temperatures the system has various ordered 
phases (spin-Peierls (SP), antiferromagnetic (AF),
spin-density wave(SDW)) and superconducting (SC)). 
The nature of the molecule (TMTTF vs TMTSF) or
of the ions slightly changes the
interchain hopping and the dimerization. Such changes 
can also be induced by applying pressure to
the system. This modifies the relative importance of 
the kinetic energy and Coulomb interation
and leads to a very rich phase diagram.
The chemical and pressure changes have similar effects, 
which can be summarised by the unified phase
diagram of Fig.~\ref{fig:unified}.

\subsection{Mott insulators and Luttinger liquids} \label{sec:transport}

\begin{figure}
\centerline{\includegraphics[width=\figwidth,height=7cm]{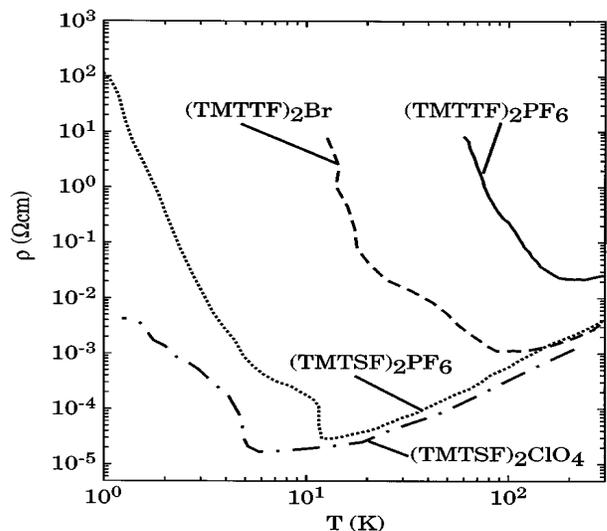}}
\caption{
Resistivity along a-axis for a series of compounds 
showing the gradual evolution to
metallic behavior as pressure is increased 
(from \cite{vescoli_photoemission_tmtsf}).}
\label{fig:resistivity_a-axis}
\end{figure}

At ambient pressure, the (TMTTF)$_2$PF$_6$ compound displays insulating
behavior (MI) .
Upon increasing pressure, a transition to a metallic
phase is found, and the properties of the TMTTF
compounds evolve toward those of the compounds of
the TMTSF family, which are good conductors.
This evolution is clear from the a-axis resistivity measurements
in Fig.~\ref{fig:resistivity_a-axis}.

The minimum of the resistivity (followed by an activated 
law as temperature is lowered) defines the onset of the
MI regime in Fig.~\ref{fig:unified}.
Such an insulating behavior in a quarter (or half filled) 
system suggests that it is
due to the interactions and that the TMTTF family is a
Mott-Hubbard insulator. It is thus clear that the 
interactions play a crucial role
in the TMTTF family even at relatively high energies. 
For the TMTSF the question is
more subtle in view of the metallic behavior at ambiant 
pressure and it was even suggested
that such compounds could be described by a FL behavior 
with weak interactions \cite{gorkov_sdw_tmtsf}.
Another important question is of course the reason for such
a difference between the very close families TMTTF and TMTSF, 
for which the various characteristics
(bandwidth, dimerization, interactions) vary relatively little.

\begin{figure}
\centerline{\includegraphics[width=\figwidth,   
clip=]{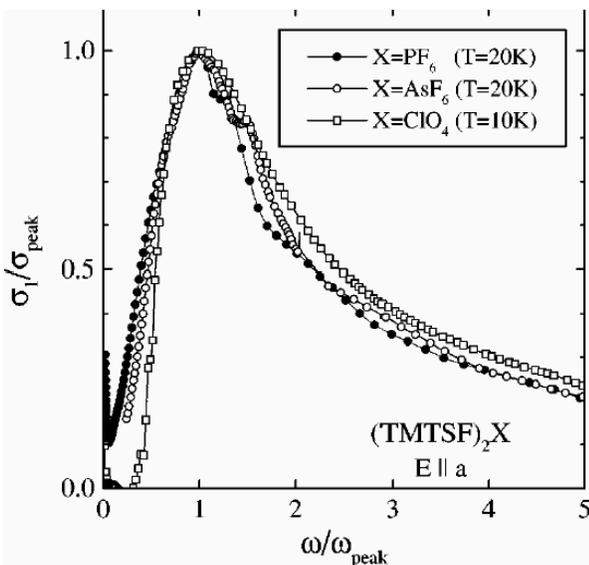}}
\caption{
Optical conductivity in the TMTSF family. Although these 
compounds seem to have a rather ``standard''
d.c. conductivity, all the d.c. transport is in fact due 
to a very narrow Drude peak containing only
1\% of the spectral weight, whereas 99\% of the spectral 
weight is above an energy gap (of the order of
$200$ cm$^{-1}$), and is reminiscent
of a Mott insulating structure. A fit of the $\omega$ 
dependence of the conductivity above the gap is
well consistent with Luttinger liquid behavior 
(from \cite{schwartz_electrodynamics}).}
\label{fig:opticala}\end{figure}

A blatant proof of the importance of interactions for {\it both}
the TF and SF compounds is provided
by the optical conductivity 
\cite{dressel_optical_tmtsf,schwartz_electrodynamics},
as shown in Fig.~\ref{fig:opticala}.
The optical conductivity clearly shows that the high 
energy structure is the one of a Mott insulator, with
a decreasing gap (of the order of 2000 cm$^{-1}$ for 
the TMTTF$_2$(PF$_6$) to 200 cm$^{-1}$ for
TMTSF$_2$(PF$_6$).
Nearly all (99\%) of the spectral weight is 
in this high energy structure.
In the metallic compounds there is in addition 
a very narrow Drude peak.
The optical conductivity shows thus clearly that 
these compounds are very far from simple Fermi liquids.
In addition one can compare the optical data with 
the theoretical predictions
\cite{giamarchi_umklapp_1d,giamarchi_mott_shortrev} 
for a LL. The data above the gap fits very well the
power law LL behavior and thus shows quite convincingly 
that these compounds are indeed well described
by a LL theory down to a scale of a few hundred Kelvin
(temperature or frequency). This is also consistent 
\cite{georges_organics_dinfiplusone}
with the optical data along the c-axis
\cite{henderson_transverse_optics_organics}, 
depicted in Fig.~\ref{fig:transverseopt}.
These measurements
directly probe the density of excited states in the $a-b$ plane.
We note however that, although clearly revealing 
that electrons are confined
in the chains above $\sim 100 K$, the measurements 
of dc transport along the c-axis
\cite{moser_conductivite_1d} (see Fig.~\ref{fig:caxis}) 
are not yet fully understood theoretically from a 
LL picture (see \cite{georges_organics_dinfiplusone} for a
discussion).

The a-axis optical measurements described above even 
allow for a quantitative determination
\cite{schwartz_electrodynamics}
of the LL parameter $K_\rho$, yielding 
$K_\rho \simeq 0.23$, indicating very strong electron
interactions
\footnote{We use the conventions of 
Ref.~\cite{schulz_houches_revue}, see also
section~\ref{sec:modelling} below. $K_\rho=1$ 
corresponds to non-interacting electrons.}.
This estimate of the LL parameter agrees reasonably 
well with measurements of the longitudinal resistivity
in the range $100 - 300K$ \cite{jerome_organic_review}. 
Photoemission data
\cite{dardel_photoemission,vescoli_photoemission_tmtsf} 
are also consistent with this value.

In addition to providing strong evidence for the 
Luttinger liquid behavior, the optical data and its
comparison to the theoretical predictions force 
one to reinvestigate the standard interpretation
for the difference between TMTTF and TMTSF compounds. 
Indeed the commonly accepted point of view
since the pioneering work of 
Ref.~\cite{emery_umklapp_dimerization} was that
the insulating behaviour of the TMTTF family
is due to the stronger dimerization of these compounds, which
effectively changes the band filling. As mentioned above, on
average two TMTSF or TMTTF molecules donate one electron to the
(monovalent) anion, so that the conduction band is nominally a
quarter filled hole band. However, the dimerization between
neighboring molecules opens a dimerization gap in the middle of
the band. When this dimerization is large, the system might best
be thought of as half-filled (TMTTF) rather than quarter-filled
(TMTSF). From this point of view the Mott insulating 
behavior comes from the half filled nature of the system.

However the optical data are inconsistent with a half filled description
in the SF family \cite{schwartz_electrodynamics}.
A more satisfactory explanation is thus to consider both series of compounds
as quarter-filled, and the Mott insulator to be due to the quarter filled
commensurability \cite{giamarchi_mott_shortrev}.
Increasing pressure or changing the chemistry reduces the 
relative strength of Coulomb interactions
with respect to hopping, hence suppressing Mott localization. For
this interpretation to be tenable, these compounds must be very
close to the Mott transition point, which is indeed supported by
the measured values of the LL parameter (indeed, $K_\rho^c=0.25$ is the
critical value for the opening of a Mott gap in a quarter-filled chain).

\subsection{Dimensional crossover and deconfinement}

Due to the interchain hopping a dimensional crossover 
will take place at low energy
between decoupled chains and a higher- dimensional behavior. Since the
isolated chains would be insulators (because $K_\rho$ is so small),
the interchain hopping can induce a deconfinement
transition provided that it becomes larger than the Mott gap. 
The system will thus
crossover from a regime where one has essentially uncoupled 
(insulating) chains to
that of metallic planes. Understanding the characteristics 
of such a transition (energy scale,
critical values of the hopping, physical nature of the 
various phases) is one of the most
challenging questions on these systems, on which we shall 
focus in the following.

\begin{figure}
\centerline{\includegraphics[width=\figwidth,height=8cm]{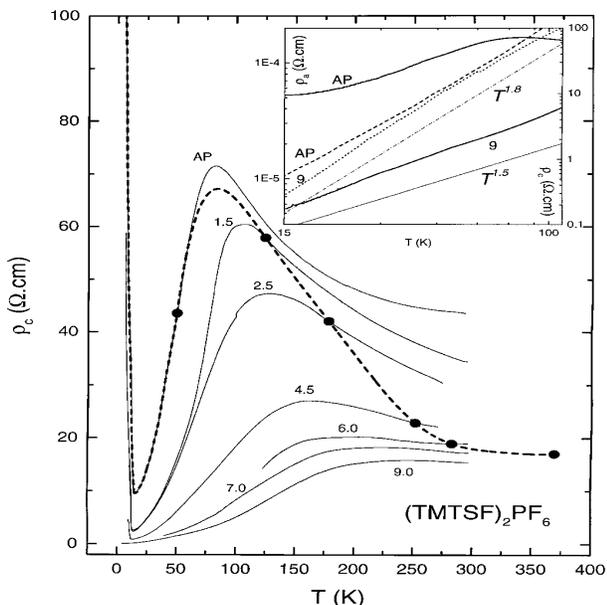}}
\caption{
Transport along the least conducting axis. This is in 
effect a tunnelling experiment between the $ab$ planes and
thus a measure of the density of states in such planes. 
In the 1d regime there are no single particle excitations and
the resistivity increases with decreasing temperature, 
whereas when the chains couple coherently one recovers a
more conventional (metallic) behavior. The maximum in 
resistivity thus measures the scale for the dimensional
crossover (from \cite{moser_conductivite_1d}).}
\label{fig:caxis}\end{figure}

\begin{figure}
\centerline{\includegraphics[width=\figwidth,clip=]{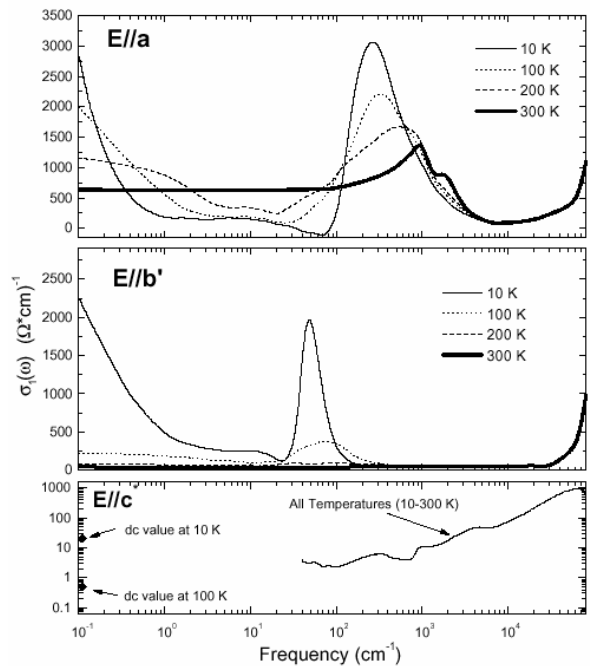}}
\caption{
Optical conductivity along the three axis, at various 
temperatures. Depending on the temperature a different
behavior is observed along the $b$ direction which 
signals a dimensional crossover from essentially uncoupled chains
to coherent planes. 
(from \cite{henderson_transverse_optics_organics}))}
\label{fig:transverseopt}
\end{figure}

A dimensional crossover is indeed observed in the transport 
along the $c$ axis shown in Fig.~\ref{fig:caxis}.
From this experiment, we see that the dimensional crossover 
takes place around $100K$ in (TMTSF)$_2$PF$_6$.
This is in agreement with the change of behavior 
from $T$ (LL behavior) to
$T^2$ observed in dc transport along the $a$ 
axis \cite{jerome_organic_review},
and with the change of
behavior in the transverse optical conductivity 
(see Fig.~\ref{fig:transverseopt}.)

\begin{figure}
\centerline{\includegraphics[width=\figwidth,height=8cm]{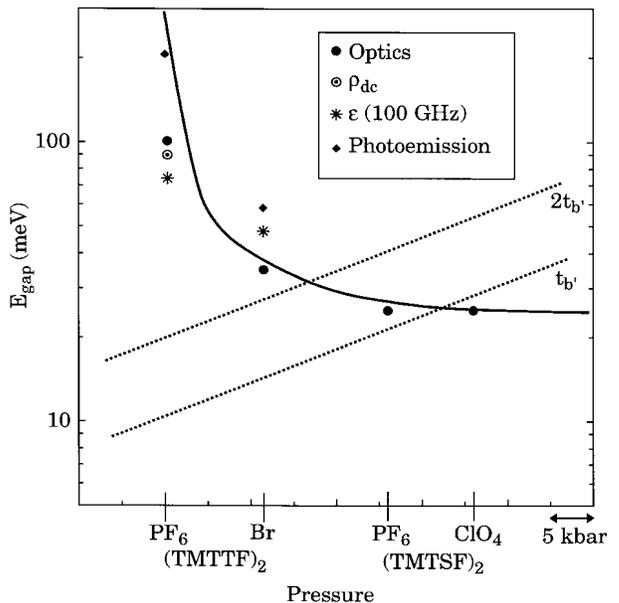}}
\caption{
A comparison of the measured gap in the optical conductivity 
with the interchain hopping. The change of
behavior from insulating to metallic occurs when the two 
quantities are of the same order of magnitude
showing that the difference between the various members 
of the TM families is indeed linked to a
deconfinement transition  
(from \cite{vescoli_photoemission_tmtsf}).}
\label{fig:confine}\end{figure}

The interpretation that the change of behavior between 
the insulating and metallic regimes is indeed due to such
deconfinement transition \cite{giamarchi_mott_shortrev} 
can be strengthened by the optical
data (see Fig.~\ref{fig:confine}).
A measure of the gap extracted from the optical
conductivity shows that the change of nature occurs when 
the observed gap is roughly of the order of magnitude
of the interchain hopping \cite{vescoli_confinement_science}.

\subsection{Summary of some open issues and questions}

To summarize the experimental situation, here are some 
key questions raised by the
physical properties of the TM family:
\begin{enumerate}
\item Are TMTSF strongly correlated systems ? What is 
the strength of interactions~?
\item What causes the difference between the various 
members of the organic families
(TMTSF and TMTTF)~?
\item Is the high temperature metallic regime a 
Luttinger liquid~?
\item At what energy scale does the dimensional 
crossover take place~?
\item What is the nature of the metallic state 
of the TMTSF series in the temperature
range from $10$ to $100K$~?
\item What are the physical properties of the low-T 
(Fermi-liquid ?)\, metallic regime~?
\end{enumerate}

As we have seen, one is now in a position to have 
satisfactory answers to questions 1-4. In doing so one
had to reexamine most of the commonly accepted point 
of views. The optical data shows that the dimerization
plays little role and that the Mott insulating behavior 
is due to the {\it quarter filled} nature of the compounds 
\cite{giamarchi_mott_shortrev,schwartz_electrodynamics}.
The estimate of the crossover scale of $\sim 100K$ 
questions early interpretations of
measurements of the NMR relaxation time \cite{bourbonnais_rmn}
$T_1$ on (TMTSF)$_2$ClO$_4$. These showed the Korringa law typical
of FL behavior $1/(T_1 T)= \mbox{const}$ at very low temperatures.
However, strong deviations from the Korringa law are observed already
around $\sim 10 K$, which is considerably
smaller than the onset of FL behavior estimated from
optics and transport.
This makes question 5) particularly puzzling,
since the NMR is anomalous in a temperature
range for which one now knows that the compounds are 
{\it not} in a one-dimensional regime.

\section{Modeling quasi-one-dimensional systems} \label{sec:modelling}

\subsection{Non-perturbative effects of inter-chain hopping}
\label{sec:non_pert}

We now turn to simplified models in which one-dimensional chains of
interacting electrons are coupled by an inter-chain hopping $\tp$.
Later on we shall concentrate on the simplest case of the Hubbard model,
but at this stage
we would like to make some remarks with a broader degree of validity.

\begin{figure}
\centerline{\includegraphics[width=\figwidth]{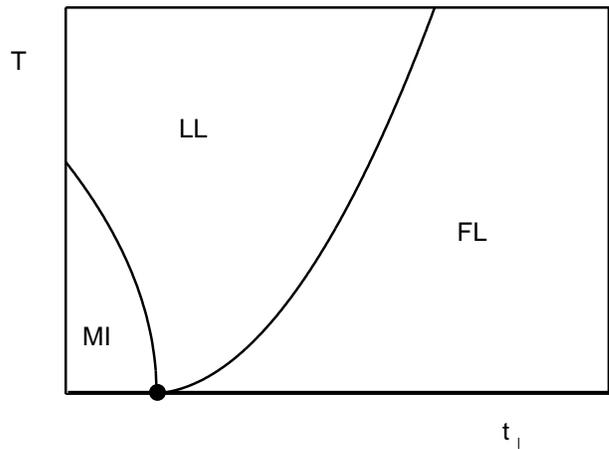}}
\caption{
Schematic representation of the crossovers in 
the commensurate case. An
insulator to metal transition occurs  for a critical 
value of $\tp$ at zero-temperature.
Crossovers are expected when reducing temperature, 
from a high-temperature Luttinger liquid behavior
to either an insulating state or a Fermi liquid metal. 
At incommensurate fillings, there is
no MI phase and only the LL to FL dimensional 
crossover line is present.
Phases with long-range order have been omitted 
on this figure.}
\label{fig:schematic_phase_diag}\end{figure}

Let us consider first the case of a commensurate filling. 
For a single chain,
strong enough interactions will then open a Mott gap, 
leading to an insulating state
in which the electrons are confined on the chains.
Technically, this is due to umklapp scattering processes 
being relevant.
This will actually happen for arbitrarily small repulsive 
interactions in the half-filled
Hubbard model, while at quarter-filling a strong 
enough nearest-neighbour interaction is
needed, in addition to a local Hubbard term.
(This is one of the reasons why a simple Hubbard 
model is insufficient to describe the
organic compounds, another one being the small 
values of $K_\rho$ needed to fit the data).
The interchain hopping will tend to delocalise 
the electrons, resulting in an insulator
to metal transition at zero-temperature for a 
critical value of the inter-chain hopping.
The generic situation is depicted on the schematic 
phase diagram of Fig.~\ref{fig:schematic_phase_diag}.

In addition to the $T=0$ insulator to metal 
(``deconfinement'') transition, finite-temperature crossovers 
have been depicted on this diagram.
For temperatures smaller than the one-dimensional bandwidth 
(i.e the high-energy cutoff of the problem), but higher 
than the crossover scales depicted in the figure,
the system is essentially
insensitive to the interchain hopping. In this regime, 
the system is expected to behave as a
Luttinger liquid (LL), which is the 
generic state for interacting
one-dimensional electrons. However, 
one should bear in mind that, at
least for small $\tp$, the Luttinger 
liquid parameter $K_\rho$ will be
gradually renormalized downwards by 
the (relevant) umklapp scattering as
temperature is reduced. The temperature 
scale at which the effective $K_\rho$
approaches zero signals the low-T Mott 
insulating regime: this is the
first crossover line depicted in the figure.
To the right of the $T=0$ deconfinement 
transition, we expect a crossover
from a high-T Luttinger liquid regime into 
a low-T Fermi liquid regime (FL).
The crossover separates a high-T regime 
in which thermal fluctuations destroy
coherent interchain transport from a low-T 
regime in which interchain
coherence sets in.
At incommensurate fillings, the situation is 
somewhat simpler: no Mott
insulating state exists, and the decoupled 
chain is always in the metallic LL
regime. Only the LL to FL crossover survives. 
Since for reasonable values of
$K_\rho$ the interchain hopping is a relevant 
perturbation
\cite{bourbonnais_couplage,wen_coupled_chains,yakovenko_manychains},
this crossover line starts
at a $T=0, \tp=0$ and the ground-state is 
a FL for arbitrarily small but finite
$\tp$.

Of course, this discussion and Fig.~\ref{fig:schematic_phase_diag}
will be complicated at very low temperatures
by the intervening of phases with long-range 
order. This will depend however
on the details of the specific model, 
in particular on the degree of deviation
from perfect nesting. We do not address 
these issues here, being interested in
compounds for which ordering temperatures 
are significantly smaller than the
crossovers described above.

It is quite tempting to compare 
the schematic crossover diagram in
Fig.~\ref{fig:schematic_phase_diag} 
to the physical properties of
quasi one-dimensional organic compounds 
described in the previous
section and summarized by the unified 
phase diagram of Fig.~\ref{fig:unified}.
Indeed, it has been advocated 
\cite{giamarchi_mott_shortrev,vescoli_confinement_science} 
that
the change from insulating to 
metallic behavior observed in
Bechgaard salts by increasing 
pressure or when going from the
TMTTF to the TMTSF family, is 
associated with a deconfinement
transition.
Obviously, increasing pressure does 
not only affect interchain hopping in
these compounds, but the deconfinement 
transition may be of a
similar nature than the one induced by 
increasing $\tp$ in a simplified
model.
The first crossover would then 
correspond to the opening
of the Mott gap (upturn of the 
resistivity in the TMTTF compounds),
while the second one would correspond 
to the onset of interchain coherence
observed in the TMTSF family.
For these reasons, the study of these 
crossovers within simple models
is of great importance for understanding 
the physics of quasi one-dimensional
organic  conductors.

The qualitative discussion above 
already hints at the fact that perturbation
theory in the interchain hopping 
$\tp$ is only of limited use for the study of
the crossovers. Let us make this 
statement more precise, and assess how far
perturbative treatments can take 
us, starting with the simpler case of an
incommensurate filling. In the 
non-interacting case, the dimensional
crossover will occur when the energy 
(e.g the thermal
energy $kT$, or the energy $\hbar\omega$ 
associated with probing the
system at a given frequency) is of the 
order of the warping of
the Fermi surface due to interchain hopping. 
Hence the crossover scale will simply
be $kT^*,\hbar\omega^*\sim \tp$. 
In the interacting case, the situation is more
complicated: the decoupled chains are 
Luttinger liquids, and $\tp$ is a
relevant perturbation. A perturbative RG 
treatment will indicate that this
effective inter-chain hopping grows as the 
energy scale is reduced: the crossover
will occur when this running coupling 
reaches $kT$. This leads to the following
estimate \cite{bourbonnais_rmn,bourbonnais_couplage} 
of the crossover scale
$\est \equiv kT^*,\hbar\omega^*$:
\begin{equation}
\label{eq:scale_RG}
\est \sim \tp (\tp/t)^{\alpha/(1-\alpha)}
\end{equation}
where $\alpha = \frac{1}{4}(K_\rho + 1/K_\rho) 
- \frac{1}{2}$ is the exponent associated
with the one-electron Green's function in the LL state. 
The important physical content
of this expression is that interactions can 
significantly reduce the crossover scale
\cite{bourbonnais_rmn,wen_coupled_chains,yakovenko_manychains},
as compared to the non-interacting estimate $\sim\tp$. We note that
(see Sec.~\ref{sec:organics_phys}) for the TMTSF 
Bechgaard salts, $\alpha$ appears to be
close to $1/2$. Since $t_b\simeq 300 \mbox{K}$ 
and $t_a/t_b\simeq 10$, this would place
$E^*$ in the 10-30 K range according to the above RG estimate.
Experimentally however, interchain coherence appears to set in
at a significantly higher temperature, 
in the 100 K range: this is one
of the puzzles in the field.

While this perturbative RG analysis allows 
to estimate a scale for the dimensional
crossover, it breaks down for $T<\est$ 
since the effective $\tp$ flows to large values.
In particular, it does not provide informations 
on the detailed nature of the low-T
Fermi liquid regime. Thus, a proper handling 
of the dimensional crossover in quasi
one-dimensional systems has to resort to 
techniques which are not perturbative in $\tp$.
This is even clearer in the case of a commensurate 
filling. If one starts from the 1D Mott insulator
fixed point, the deconfinement transition 
is clearly a non-perturbative phenomenon since the
inter-chain hopping is an irrelevant perturbation 
at this fixed point. If, on the other hand, one
starts with the LL fixed point associated with 
the high-T regime, then one has to deal
simultaneously with {\it two} relevant perturbations: 
the umklapp scattering (responsible
for Mott physics) and the interchain hopping. 
The competition between these two relevant
perturbations will determine the crossovers described above.

Non-perturbative studies are thus needed to 
investigate both the deconfinement transition and
the dimensional crossover. Several authors 
have developed such methods in the case of
{\it a finite number} of coupled chains.
Below, we present a recently developed 
approach designed to handle
{\it an infinite array of coupled chains} 
in a non-perturbative 
manner\cite{arrigoni_tperp_resummation,%
georges_organics_dinfiplusone,biermann_dmft1d_hubbard_short}.

\subsection{The chain- Dynamical Mean Field Theory approach}

This non-perturbative method is inspired by the success of
dynamical mean field theory
(DMFT) in the description of lattice models of
correlated fermions~\cite{georges_d=infini}.
Ordinary DMFT maps a lattice model of
interacting fermions onto a single site model in an
effective time-dependent mean field, which has to be
determined self-consistently. Thus the problem becomes
equivalent to an Anderson impurity model in a
self-consistent bath. It can be shown that
in the limit of an infinite coordination number of
the lattice, all quantum fluctuations become local
and the dynamical mean field description is exact.

Very anisotropic systems of coupled one-dimensional chains
lend themselves to a very natural extension of this approach
(dubbed ``chain-DMFT''),
in which the array of chains is replaced by an effective chain in a
self-consistent bath. It is crucial to retain a one-dimensional
geometry of the effective problem, if the limit of decoupled chains
(and hence 1D physics) is to be treated properly. 
The self-consistent bath describes the influence of
all neighboring chains on the "impurity chain", thus
freezing all quantum fluctuations that are non-local
in the transverse direction, but retaining spatial and dynamical
fluctuations along the chains. By the same arguments
as in ordinary DMFT, the chain-DMFT approach becomes
exact in the limit of an infinite transverse
coordination number of the array. In the limit of decoupled
chains chain-DMFT reduces to the single chain
problem, which is solved numerically exactly.

Consider a system of coupled chains described by the
Hamiltonian
\begin{equation}
\label{ham1} H = \sum_{m}\,H_{1D}^{(m)} -
\sum_{\langle m,m'\rangle}\tp^{mm'} \, \sum_{i\sigma}
(c^+_{im\sigma}c_{im'\sigma} + \mbox{h.c})
\end{equation}
where $H_{1D}^{(m)}$ is the Hamiltonian for a single
isolated chain and the sum in the second term runs
over neighboring chains $m,m^{\prime}$, which are coupled by the
transverse hopping $\tp^{mm'}$.
Chain-DMFT maps this system onto a single-chain system
described by the effective action
\begin{eqnarray}\label{Seff}
\nonumber S_{\rm eff}=&& - \int\int^{\beta}_{0}d\tau\,d\tau'
\sum_{ij,\sigma} c^{+}_{i\sigma}(\tau) {\cal
G}_0^{-1}(i-j,\tau-\tau')
c_{j\sigma}(\tau')\\
&&+\int_0^{\beta} d\tau
H^{int}_{1D}[\{c_{i\sigma},c^+_{i\sigma}\}]
\end{eqnarray}
where $H^{int}_{1D}$ is the interacting part of the on-chain
Hamiltonian.
${\cal G}_0$ is the {\it
effective propagator} describing hopping processes, that
destroy a fermion on site $j$ at time $\tau^{\prime}$ and
create one at site $i$ at time $\tau$.
In analogy to usual DMFT ${\cal G}_0$ has to be determined
from a self-consistency condition that imposes that the Green's
function $G(i-j,\tau-\tau')\equiv -\langle c(i,\tau)c^+(j,\tau')
\rangle_{\rm eff}$ calculated from $S_{\rm eff}$ coincides
with the on-chain Green's function of the original problem. Since the
self-energy is $\Sigma={\cal G}_0^{-1}-G^{-1}$, this condition reads:
\begin{equation}
G(k,\iomn) = \int d\ep
{{D(\ep)}\over{\iomn+\mu-\epsilon_k-\Sigma(k,\iomn)-\ep}}
\label{sc_cond}
\end{equation}
where $\ep(\kp)$
denotes the Fourier transform of the
inter-chain hopping $\tp^{mm'}$,
$D(\ep)=\sum_{k\perp}\delta[\ep-\ep(\kp)]$ the corresponding
density of states,
$k$ the momentum in chain direction and $\omega_n$ the
Matsubara frequencies.
The chain-DMFT equations (\ref{Seff},\ref{sc_cond}) fully
determine the self-energy and Green's function of the coupled
chains. In particular, once $\Sigma$ has been calculated, the full
Green's function for the coupled chains is obtained from:
$G(k,\kp,\iomn)^{-1}=\iomn+\mu-
\epsilon_k-\epsilon_\perp(\kp)-\Sigma(k,\iomn)$.
We note that a key approximation of this approach 
is that the self-energy is considered to be independent 
of the transverse component of the momentum.
In the following we specialize the discussion to
a model of Hubbard chains coupled by a perpendicular
hopping $\tp$. Then the Hamiltonian is of the form
(\ref{ham1}) with $\tp^{mm'}=\tp$ and
$H_{1D}^{(m)}$ being the Hamiltonian of the one-dimensional
Hubbard model.

Different choices of the lattice
are possible. Choosing a semi-circular density of states
\begin{equation}
D(\epsilon) = \frac{1}{2 \pi t^2} \sqrt{\epsilon^2-4 t^2}
\end{equation}
in Eq.~(\ref{sc_cond}) corresponds to a Bethe lattice
with infinite coordination number,
in which each lattice site is replaced by a one-dimensional
chain and hopping between the chains follows the topology
of the Bethe lattice.
This choice is particularly appealing, since the chain-DMFT
formalism is exact in this case.

Another possibility is to consider a two dimensional array of chains
with the square lattice geometry. In this case,
each chain has exactly two nearest neighbors and chain-DMFT 
must be viewed as an approximation. 
The transverse dispersion becomes:
\begin{equation}
\epsilon_\perp(\kp)\,=-2\tp\,\cos\kp
\end{equation}
and the corresponding density of states is the one of a one-dimensional
lattice:
\begin{equation}
D(\epsilon) = \frac{1}{\pi} \frac{1}{\sqrt{\epsilon^2-4t^2}}.
\end{equation}
As for the Bethe lattice, the self-consistency condition
simplifies in the 2D geometry in such a way that the Weiss field
can directly be written in terms of the Green's function:
\begin{equation}
{\cal G}_0^{-1}(k,\iomn)
=\iomn + \mu + G^{-1}(k,\iomn)
- \sqrt{G^{-2}(k,\iomn) + 4 \tp^2}.
\end{equation}
This model does not provide a controlled limit, in which chain-DMFT
becomes exact. Still, from a conceptual point of view, a two-dimensional
picture seems more appealing if one aims at a comparison with
the experimental situation in the Bechgaard salts.
Therefore the calculations presented below were performed for
this choice of the model.
We stress, however that in practice, the specific choice of the
transverse dispersion does not have a significant qualitative influence
(as long as one does not address long-range ordering).

A practical implementation of the chain-DMFT approach requires to
solve the effective one-dimensional interacting problem described above.
This is a rather formidable task, and numerical methods are required.
Even though other techniques are conceivable,
we have chosen to use a Quantum Monte Carlo algorithm, which
is a straightforward generalization of the
Hirsch-Fye algorithm used in single-site DMFT
\cite{hirsch_fye_qmc,georges_d=infini}.
It relies on a Trotter discretization of the effective action
in imaginary time, and on a discrete Hirsch transformation
of the interaction term:
\begin{eqnarray}\label{hir}
e^{- \Delta \tau U n_{i \uparrow} n_{i \downarrow}}
= \frac{1}{2} \sum_{s=\pm 1}
e^{\lambda s \left( n_{i \uparrow}-  n_{i \downarrow} \right)
- \frac{\Delta \tau U}{2}
\left( n_{i \uparrow}+  n_{i \downarrow} \right)}
\end{eqnarray}
with 
$\lambda=\mbox{arccosh}\left( e^{\frac{\Delta \tau U}{2}}\right)$. 
An Ising field $s$ is introduced
at each time slice and each site of the chain.
Monte Carlo sampling of the Ising fields then allows for
the direct calculation of the on-chain Green's function and
on-chain correlation functions.
In practice, chains of 16 to 32 sites
with periodic boundary conditions are sufficient to
access the 1d Luttinger Liquid regime.

In the following we will present numerical evidence for
the deconfinement transition at half filling and the dimensional
crossover as a function of temperature in the doped case.

\subsection{Luttinger-liquid to Fermi-liquid crossover}

We summarize some of our QMC results for an array of
coupled Hubbard chains in the chain-DMFT approach, starting
with the doped (incommensurate) case.
At high temperature, the model is expected to display
LL behaviour. In order to measure numerically the LL parameter
$K_\rho$, we have computed the local
spin-spin correlation:
\begin{equation}
\chi_s(\tau)=\langle
S^z(j,0)S^z(j,\tau)\rangle =
\sum_{k,k\perp}\chi_s(k,k_{\perp},\tau)
\end{equation}
In a LL, the asymptotic behavior of this quantity
reads, in imaginary time:
\begin{equation}
\chi_s(\tau)=\chi_s(\beta/2)\,
\left(\frac{1}{\sin\pi\tau/\beta}\right)^{1+K_\rho},
\end{equation}
valid for $\beta$, $\tau$ and $\beta-\tau$ larger than the inverse of the
high-energy cutoff (i.e in some range around $\tau=\beta/2$).
We emphasize that this is a much better manner of accessing $K_\rho$
in a QMC simulation than by looking at the single-electron Green's function,
whose asymptotics involves the exponent $\alpha=(K_\rho+1/K_\rho)/4-1/2$
which is never very large for the Hubbard model. This makes
it very hard numerically to distinguish LL from FL behavior at the
level of one-electron
Green's functions.
\begin{figure}
\centerline{\includegraphics[width=\figwidth]{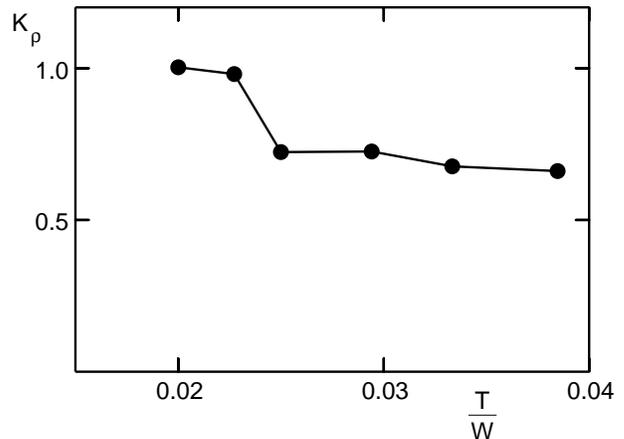}}
\caption{ Effective $K_\rho$ {\it vs.} temperature
in the doped case (filling $n \simeq 0.8$) for $U/W=1.0$,
$\tp/W=0.14$.}
\label{fig:kro_vsT_doped}\end{figure}

In Fig.~\ref{fig:kro_vsT_doped}, the effective value of $K_\rho$ obtained by
fitting the local spin correlation function to the above asymptotic form is
plotted as a function of temperature
\footnote{We give all energies in units of the longitudinal
band width $W=4t$.}, for $U/W=1$, a filling $n \simeq 0.8$ and $\tp/W=0.14$.
This plot clearly reveals the expected dimensional crossover from a
high-T LL regime with $K_\rho<1$ to a low-T FL regime with $K_\rho=1$.
The numerically measured value $K_\rho\simeq 0.7$ in the LL regime is
in very good agreement with the exact value known
from analytical (Bethe ansatz) calculations on the
one-dimensional Hubbard model, for that value of 
$U/W$ \cite{schulz_conductivite_1d}.

Our results for the crossover scale are consistent
with $T^* \simeq C t_\perp/\pi$ with $C\simeq 0.5$.
We cannot meaningfully test the RG prediction
$T^* = \frac{\tp}{\pi}C (\tp/t)^{\alpha/(1-\alpha)}$
for the
reduction of this scale due to interactions
\cite{bourbonnais_rmn,bourbonnais_couplage}, because of the
very small value of $\alpha\simeq 0.03$ for the
present model. As discussed in the last section, extensions
of our calculations to models with smaller values of $K_\rho$ are
required both for their theoretical interest and for a meaningful
comparison to organics.

\subsection{Deconfinement transition}

\begin{figure}
\centerline{\includegraphics[width=\figwidth]{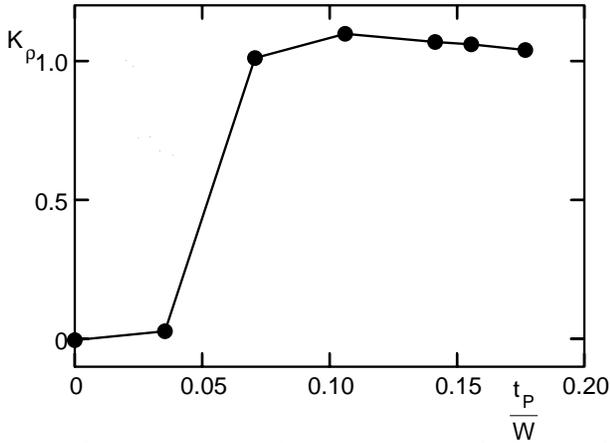}}
\caption{
$K_\rho$ as a function of the perpendicular hopping $t_\perp$ 
for a half filled
system.}
\label{fig:kro_vs_tp_half}\end{figure}

We now turn to the half-filled case. In Fig.\ref{fig:kro_vs_tp_half}, we
display the effective $K_\rho$ (determined as above) as a
function of interchain hopping, for $U/W=0.65$ and
at a rather low temperature $T/W=0.025$.
For small $\tp/W$, the value $K_\rho=0$ indicating a Mott insulating
behaviour (with a decay of the spin-spin correlation similar to
that of a Heisenberg spin chain). In that regime, the calculated charge
correlation function (not shown) clearly displays the exponential
decay associated with a finite charge gap.
Beyond a critical value of $\tp/W$, we find $K_\rho\simeq 1$,
signalling a FL regime (and a corresponding behavior for the
charge correlation function). Hence, the expected deconfinement
transition (Fig.\ref{fig:schematic_phase_diag}) is clearly
revealed by our calculations.
In principle, it should be possible to identify first
a Luttinger Liquid phase and then, with increasing
interchain coupling, a Fermi Liquid phase.
However, in this parameter range the Luttinger Liquid
phase is too narrow to become visible.
The location of the deconfinement transition is in
reasonable agreement with the naive criterion $\Delta_{1D} \sim
t_\perp^{eff}$, with $\tp^{eff}$ the renormalised inter-chain hopping.
\cite{giamarchi_mott_shortrev,vescoli_confinement_science,%
tsuchiizu_confinement_spinful_refs}.

\subsection{Interchain optical conductivity}
Inter-chain optical conductivities within
chain-DMFT can be obtained from the one-particle
Green's functions. Vertex corrections
drop out for analogous reasons as in single-site DMFT, and
therefore:
\begin{eqnarray}
\nonumber \R\,\sp (\o,T) \propto &&\tp^2\int {{d\kp}\over{2\pi}}\,
\sin^2\kp\,
\int {{dk}\over{2\pi}}\int d\omega'A(\ep,k,\o')\\
&&\times A(\ep,k,\o+\o'){{f(\o')-f(\o'+\o)}\over{\o}}
\label{sigmatrue}
\end{eqnarray}
where $A(\ep,k,\o)=-{1\over\pi}\mbox{Im}G(\ep,k,\o)$ is the
single-particle spectral function of the coupled chains system.
Note that we have taken into account the $\kp$- dependence of the
current vertex in this formula.
\begin{figure}
\centerline{\includegraphics[width=\figwidth]{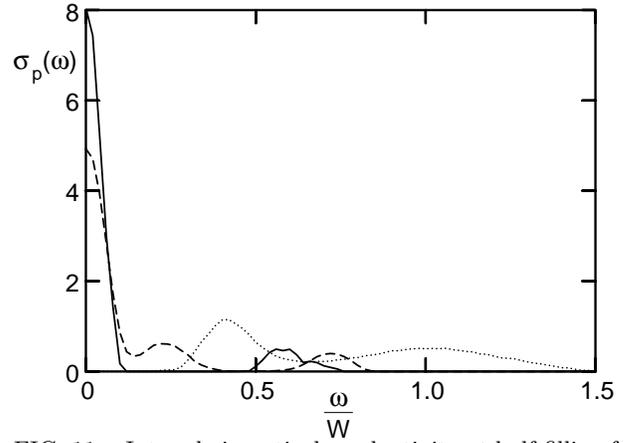}}
\caption{ Inter-chain optical conductivity at
half-filling for $U=0.65 W$, $\beta=40/W$ and
$\tp=0.14$ and $0.07$ (solid and dashed lines) and
for $U=1.0W$, $\beta=40/W$ and $\tp=0.$ (dotted line).}
\label{fig-conduc}\end{figure}
Fig.~\ref{fig-conduc} shows the inter-chain optical conductivity at
half-filling for $U=0.65 W$, $\beta=40/W$ and
$\tp=0.14$ and $0.07$ (solid and dashed lines) and
for $U=1.0W$, $\beta=40/W$ and $\tp=0.$ (dotted line).
As expected, in the case of uncoupled chains ($\tp=0.$)
the system exhibits a Mott gap, followed by an onset of
absorption starting at approximately the gap and
extending up to a scale of order U, where a broad second
peak is apparent. As the insulator to metal transition
is crossed, a low-frequency Drude peak develops.
Close to the transition the weight of the Drude peak
is small, while the Hubbard band feature is still visible and
carries a significant part of the spectral weight.

To make comparison to the Bechgaard salts, where the
Drude peak was shown to carry less than $2\%$ of the
spectral weight with $98\%$ of the weight being contained
in the Hubbard features, more realistic models should
be considered. In particular, those should allow
for stronger effects of the
interactions, that is smaller values of $K_\rho$.
We note, however, that the
general tendency of coexistence of a small Drude
peak with strong Hubbard bands is already visible
in the Hubbard model.

\subsection{The low-temperature Fermi Liquid regime}

\begin{figure}
\centerline{\includegraphics[width=\figwidth]{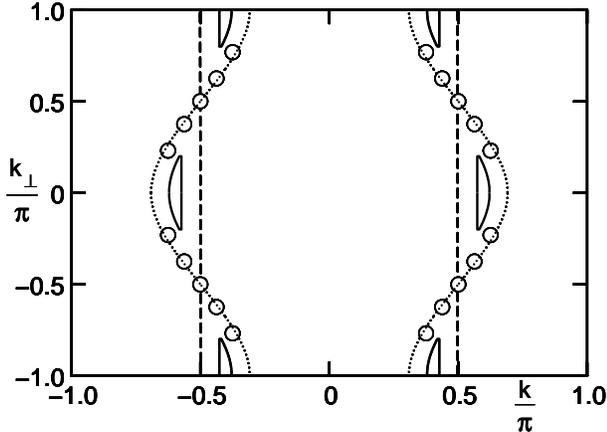}}
\caption{ FS in the half-filled case
with $\tp/W=0.14$, $U/W=0.65$ (circles), compared
to the FS of the non-interacting case (dotted line) and of the
purely 1d case ($\tp=0$ -dashed-.). The solid line depicts
schematically the FS obtained within the
RPA ($\Sigma=\Sigma_{1D}$) \cite{essler_rpa_quasi1d}.}
\label{FS-plot}\end{figure}

At strong enough transverse coupling the system becomes
a FL. In our numerical simulations, the onset of the FL regime is
identified from the behavior of $K_\rho$ (see Fig.~\ref{fig:kro_vs_tp_half})
and from a linear behavior of the imaginary part of the
self-energy in Matsubara space:
$\Sigma(k, i \omega) \sim i \omega$.
The equation defining the Fermi surface
$\mu-\epsilon_k-\Sigma(k,0)-\ep=0$
then yields a relation $\kp(k)$ for the points $(k,\kp)$
that lie on the Fermi surface.
These are visualized in Fig.~\ref{FS-plot} for the half-filled case.
For the uncoupled (1d) system the Fermi surface consists of
straight lines (dashed lines in the figure); the transverse
hopping induces some cosine-like modulation
but does not change the topology drastically.
Indeed, the Fermi surface of the interacting coupled
system (circles in Fig.~\ref{FS-plot})
is very close to the one of non-interacting ($U=0$.)
coupled chains (dotted line in Fig.~\ref{FS-plot}).

This is in striking contrast with predictions of
an RPA approach, as pointed out in \cite{essler_rpa_quasi1d}.
The RPA consists of replacing the self-energy of
an interacting coupled system
by the self-energy of a 1d chain. Since the
1d chain is a Mott insulator at half-filling,
$\Sigma$ diverges at low frequency for $k=\pi/2$
and the above equation has no solution for this
$k$-value. Thus, within the RPA the Fermi surface
cannot cross the points $(\pm \pi/2,\pm \pi/2)$
but consists of disconnected
pockets as depicted schematically in Fig.~\ref{FS-plot}
(solid lines).
The feedback of the effects
of the interchain hopping on the self-energy,
which is taken into account by chain-DMFT,
regularizes the behavior of the self-energy
near these points, leading to an open Fermi surface.

The QP residue $Z_{\kp}$ calculated within chain-DMFT
depends only very weakly on the
Fermi surface point (Table I).   
\begin{table}
\caption{QP weights $Z(\kp)$ for different points on the
FS (half-filled case, $\tp=0.14 W$, $U/W=0.65$).
\label{tab1half}}
\begin{tabular}{llllll}
$\kp/\pi$ & 0.23 & 0.38 & 0.50 & 0.62 & 0.77 \\
\hline
$Z(\kp)$  & 0.79  & 0.77 & 0.76 & 0.77 & 0.79
\end{tabular}
\end{table}
Again, this shows that the regularization by interchain
hopping feedback effects is very efficient: a nearly divergent
self-energy as could be imagined to result from an only slightly
regularized 1d self-energy would lead to very small $Z(\kp)$
close to $\pm \pi/2$. Therefore approximations based on the
1D self-energy are likely to predict ``hot spots'' at those
FS points corresponding to a vanishing inter-chain kinetic energy.
Our results, in agreement with those of Arrigoni
\cite{arrigoni_tperp_resummation_prb},
do not support such a picture. Rather,
the $\kp$-dependence of $Z(\kp)$ is very weak,
with very shallow minima at $\kp\sim \pm \pi/2$.
This small variation is however on the scale of our error bars.

\section{Outlook}

In these lecture notes, we have reviewed a few of the fascinating
physical properties of quasi one-dimensional organic conductors.
Our point of view has been that a unified description of both the TMTTF and
TMTSF compounds is possible, as strongly interacting {\it quarter-filled}
chains with weak interchain couplings.
In that picture, the physical changes under pressure (or from TMTTF
to TMTSF) are viewed as a deconfinement transition from a Mott insulating
regime to a metallic regime
\cite{giamarchi_mott_shortrev,vescoli_confinement_science}.
Still, we have emphasized that it is hardly
possible to understand these compounds using a purely one-dimensional
picture. Inter-chain coherence sets in at low energy 
in the metallic compounds,
and this leaves us with a quite difficult theoretical problem, 
which cannot be
handled by perturbative techniques in the inter-chain hopping.

We have reviewed the recently developed chain-DMFT approach which is in our
opinion a promising route to handle these issues. QMC calculations on
weakly coupled Hubbard chains have clearly demonstrated that this method is
able to reproduce both the deconfinement transition and the dimensional
crossover from a LL to a FL as the energy scale is reduced.
We have also emphasized that a purely local Hubbard interaction is not
appropriate
for modelling quasi one-dimensional organics. 
Indeed, the phenomenology requires
i) the possibility of a Mott insulating state at quarter filling and 
ii) rather small values of $K_\rho$ ($\simeq 0.23$) which cannot be 
reached in the Hubbard model.
A minimal theoretical model obeying these requirements is that of weakly
coupled quarter-filled chains with both an on-site and a nearest-neighbour
interaction. We hope to be able to deal with this model in the chain-DMFT
framework in the near future.

Many outstanding questions on the physics of organics remain
unanswered at this stage. We have tried to emphasize some of these
puzzles in this paper.


\section{Acknowledgements}
We acknowledge many discussions with D. Jerome, as well
as P. Auban-Senzier, J. Moser, P. Wzietek and C. Pasquier, 
and with L. Degiorgi and G. Gr\"uner
on the physics of quasi one-dimensional organic conductors.
This research is partially supported by a Marie Curie Fellowship
of the European Community Programm ``Improving Human
Potential'' under contract number MCFI-2000804 and a
grant of supercomputing time at NIC J{\"u}lich.





\end{document}